# Rhodium Mössbauer Supperadiance Induced by Liquid-Nitrogen Cooling

Yao Cheng, Bing Xia, Zhongming Wang

Department of Engineering Physics, Tsinghua University, Beijing 100084, P. R. China

Email: yao@tsinghua.edu.cn

In the previous report, we have demonstrated cascade branching channels of the multipolar E3 transition of rhodium Mössbauer γ via the time- and energy-resolved spectroscopy. Moreover, superradiance in the Borrmann channel from inverted nuclei gives γ entanglement. In this letter, we report further four observations of superradiance and its associated γ entanglement at the liquid-nitrogen temperature, i.e. (i) speed-up decay, (ii) immediate recovery of the speed-up decay after quenching, (iii) simultaneous suppression of γ and K lines, and (iv) enhanced multiple ionizations. Anisotropic superradiant channels open by quenching and recover back immediately after quenching. Enhanced K satellites and K hypersatellites induced by cooling are attributed to the inelastic scattering of more than three entangled γs.

PACS numbers: 33.45.+x, 76.80.+y, 23.20-g, 27.60+j, 78.70.Ck

Recently, we have reported the observation of long-lived rhodium Mössbauer Effect generated by bremsstrahlung irradiation [1]. In this letter, the Mössbauer Effect with extremely narrow linewidth of $10^{-19}$ eV is further demonstrated by cooling down the sample temperature. Discovery of broad-band cascade γs [2] advance our understanding of puzzling tri-photon effect reported in [1]. More than half γ radiation goes to these cascade branching channels. The tri-photon pile-ups between 60 keV and 80 keV featured by the triple fast decay reported in [1] are thus revealed by three parts, i.e. (i) bunching K lines, (ii) pile-ups between cascade γs and K lines by accident and (iii) characteristic K lines of the high-Z impurities.

Superradiance in the Borrmann channel leads to directional entanglement of γs, which depends on temperature and geometry [2]. In this letter, we shall demonstrate that directional superradiance in polycrystal at liquid-nitrogen temperature gives multiple entangled γs, particularly three γs denoted by tri-γ. Each γ propagates like the multi-beam nuclear Borrmann mode [3]. Borrmann mode is the known x-ray eigenmode in crystal from field cancellation at the lattice sites, and thus the photo-electric effect is suppressed [4]. In the mean time, nuclear coupling can be enhanced for the multipolar nuclear transition [3]. Matching condition of the crystal lattice in the weakly attenuated Borrmann channels provides the cooperative emissions of multiple γs with correlated directionality, polarization, and phase. The cooperative emissions are not able to be factorized into single-particle states [2], which is similar to the entangled biphoton state generated by spontaneous parametric down conversion [5]. Since the Borrmann effect depends on temperature, it becomes more significant at low temperature due to reduced atomic vibration. When atoms deviate from their equilibrium positions, they become scattering centers of entangled γs and leads to bunching K lines. Consequently, the impurities, e.g. bismuth and platinum, are also scattering centers of the entangled γs.

We have applied 120-minute irradiation with the same excitation procedure reported in [1]. Measurements are carried out with the leveled HPGe detector of Fig. 2b in [1]. A small HPGe detector of CANBERRA GL0210P with an active area of ϕ-1.6 cm is applied in the N-S detecting direction. Detecting system consists of the optic feedback pre-amplifier CANBERRA 2008 BSL, the multichannel analyzer ORTEC 917A and the ORTEC 572 amplifier tuned at 3-µs shaping time. Collecting angle is about 0.4π sr. The square sample has a dimension of 2.5 cm × 2.5 cm × 1 mm with 99.9% purity of rhodium (Rh00300, Goodfellow). The rhodium sample is a polycrystal with the fcc lattice structure. When liquid nitrogen immerses the sample from one side, liquid nitrogen is able to overflow the sample such that double-side cooling is created. Typical temperature behavior is measured later as shown in Fig. 1.

Fig. 2 illustrates the spectrum accumulated in three hours. We list the assignments of significant peaks in Table 1. Kα internal conversions at 65.1 keV and 66.8 keV and γ at 98.8 keV of the 4.02-day $^{195m}$Pt transition are identified, which is impurity excited by the bremsstrahlung irradiation. The initial count rate is about 700 cps mainly contributed by K lines. Fig. 3 illustrates the time evolutions of two K lines, γ and γ escape. Speed-up decay during the cooling period and immediate recovery after quenching stop collectively demonstrate the opening and the closing of superradiant channels.

We have calibrated the γ-escape intensity as a function of γ energy by two γ sources of $^{241}$Am and $^{109}$Cd. The intensity ratio between 30-keV γ escape and 40-keV γ shall be 0.014 and not depending on sample temperature and external filter. Low-energy γ is absorbed at detector surface that gives high escape. Large ratio is revealed by reduced counts when lowering temperature in Fig. 3. The same result of large γ escape is also confirmed by differential measurements of filters in [2]. When entangled cascade γs of fixed sum energy arrive detector, the same γ energy is reproduced but with enhanced γ escape without energy broadening [2].

Fig. 4 explores the time evolutions of spectral deformations $\tilde{S}_i(\omega,t)$ deviated from the calibrated normal profiles $\bar{S}_i(\omega)$ [2]. Measured spectra $S_i(\omega,t)$ are normalized by individual total counts, channel by channel in time

$$\tilde{S}_i(\omega,t) = A_i \left( \frac{S_i(\omega,t)}{\int S_i(\omega,t)d\omega} - \overline{S}_i(\omega) \right) \quad (1)$$

with factors $A_i$ chosen for each spectrum in Fig. 4. The low index $i$ stands for Kα and Kβ and γ. Energy shifts of rhodium K satellites and K hypersatellites are about 50 eV and 500 eV due to coincident creation of two holes, i.e. one K hole + one L hole and two K holes respectively [6]. K satellites and K hypersatellites are attributed to the inelastic scattering of entangled γs, which are enhanced by an order of magnitude at low temperature (see Fig. 4). During the cooling period, strong satellite shifts are revealed by zero crossing lines of eq. (1) locating at 20.2 keV of Kα and at 22.7 keV of Kβ respectively. When temperature slowly recovers back, spectral deformation is similar to the H1 phase reported in [2] featured by hypersatellite lines. Overall, hypersatellite shift of Kβ is stronger than hypersatellite shift of Kα, while Kα$_2$ is enhanced together with Kα hypersatellites, except strong satellite shifts during the cooling period.

Three L$_1$ holes are possible, whereas three L$_2$ holes are impossible due to the fact that L$_2$ shell has only two electrons. In the case of two K holes + one L hole, Kα hypersatellite is emitted first, which gives states of two L holes + one K hole. When subsequent second photon emits, the possibility to emit Kα$_1$ satellite is higher than the possibility to emit Kα$_2$ satellite. High Kα$_1$ satellite shift reveals coincident observation of hypersatellites and enhanced Kα$_2$. Kβ hypersatellites are generally stronger than Kα hypersatellites, which also indicate coincident creation of two K holes + multiple L holes.

In the left hand side of the Kα around 19 keV in Fig. 4, a peak emerges ascribed to the cascade γs during cooling period. Figs. 5 and 6 illustrate the time evolutions of cascade γs in two separated regions, i.e. >20 keV and <20 keV. Counts in both regions follow the γ sag at 40 keV in time (see Fig.

3). Figs. 7 illustrate the ratios of Kα/Kβ and K/γ to demonstrate the negligible nitrogen absorption and stronger K suppression than γ suppression. Stronger K suppression reveals the coincident detection of entangled cascade γs back to 40 keV, otherwise the cascade branching shall give the same amount of suppression on K lines and γ.

In conclusion, speed-up rhodium decay at low temperature is a further evidence for the long-lived Mössbauer effect. Immediate recovery after quenching stop reveals that superradiance channels open due to quenching but not toward detector. Directional superradiance in favor of the long sample axis will be reported in the near future. Suppressed γ and K lines during the cooling period collectively reveal the anisotropic superradiance of branching cascade γs. Two observations of (i) coincidence of hypersatellites and $Kα_2$ enhancement; and (ii) more Kβ hypersatellites than Kα hypersatellites, together reject the entanglement merely of biphoton but tri-γ or more than three entangled γs. Entanglement of γs is attributed to the cooperative anomalous emission from the active internal Mössbauer source of E3 multipolarity in the Borrmann channel [2,7], which is enhanced by lowering the sample temperature.

We thank Hong-Fei Wang for the help of manuscript preparation, Yanhua Shi, Yexi He and Yinong Liu for fruitful discussion and Yuzheng Lin with his accelerator team, especially Qingxiu Jin and Xiaokui Tao. This work is supported by the NSFC grant 10675068.

| Rhodium | Kα escape | Kβ escape | $Kα_{1,2}$ | $Kβ_{1,2,3}$ | γ escape | γ | Kα+Kβ | Kβ+Kβ |
|---|---|---|---|---|---|---|---|---|
| Energy | ∼10 keV | ∼12 keV | 20.2 keV | 22.7 keV | ∼30 keV | 39.8 keV | 42.9 keV | 45.4 keV |
| I & S | $^{195m}$Pt $Kα_2$ | $^{195m}$Pt $Kα_1$ | Pb $Kα_2$ | Pb $Kα_1$ | $^{195m}$Pt $Kβ_1$ | Pb $Kβ_1$ | $^{195m}$Pt γ | |
| Energy | 65.1 keV | 66.8 keV | 72.8 keV | 75.0 keV | 75.7 keV | 84.9 keV | 98.8 keV | |

Table 1. Assignments of significant peaks in Fig. 2. Abbreviations of I & S, Kα+Kβ and Kβ+Kβ stand for impurity & shielding, peak pile-up of Kα+Kβ and peak pile-up of Kα+Kβ respectively.

[1] Y. Cheng, B Xia, Y.-N. Liu, Q.-X. Jin, Chin. Phys. Lett. **23**, 2530 (2005); Y. Cheng, B. Xia, J. Li, Chin. Phys. Lett. **23**, 826 (2006); **23**, 2348 (2006) [erratum]; Y. Cheng, *et al.*, Hyperfine Interactions, **167** 833 (2006)

[2] B. Xia and Y. Cheng, arXiv:0706.2628, submitted to PRL (2007)

[3] J. T. Hutton, J. P. Hannon and G. T. Trammell, Phys. Rev. A **37**, 4269 (1988).

[4] B. W. Batterman, H. Cole, Rev. Mod. Phys. **36** 681 (1964).

[5] Y. H. Shih, IEEE Journal of selected Topics in Quantum Electronics **9**, 1455 (2003).

[6] B. Boschung, *et al.*, Phys. Rev. A **51**, 3650 (1995); I. Török, T. Papp, S. Raman, NIMB **150**, 8 (1999).

[7] J. P. Hannon and G. T. Trammell, Hyperfine Interactions, **123/4**, 127 (1999); J. P. Hannon, N. J. Carron and G. T. Trammell, Phys. Rev. B **9**, 2810 (1974).

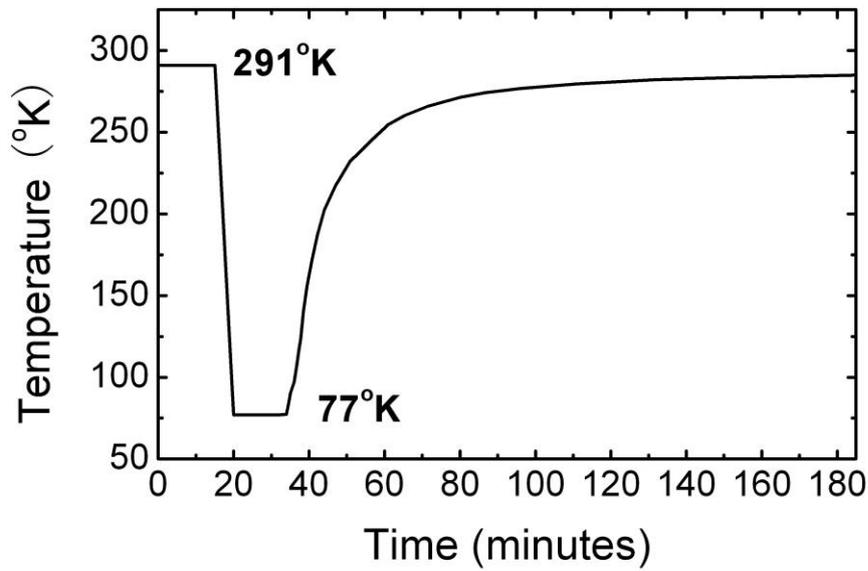

FIG. 1. Typical time evolution of the sample temperature measured with Pt 1000. The sample is kept at room temperature of 18°C for twenty minutes and then is immersed in liquid nitrogen for twelve minutes. Afterwards, sample temperature recovers slowly back to room temperature.

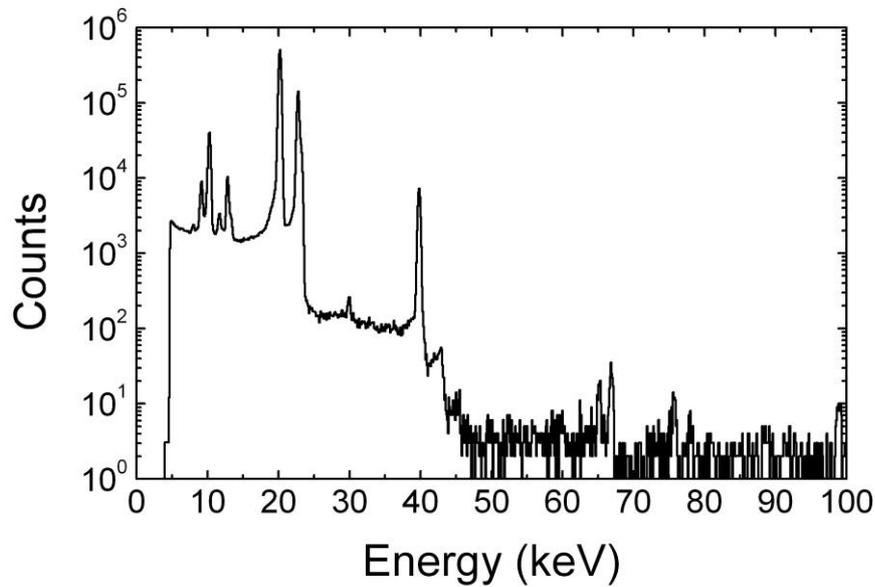

FIG. 2. Energy-resolved spectrum of rhodium isomeric emissions collected in 4096 channels for 415 keV within three hours. Energies of the significant peaks are listed in Table 1. Escape γ peak at 30 keV has enhanced intensity [2], of which time evolution is shown in Fig. 3.

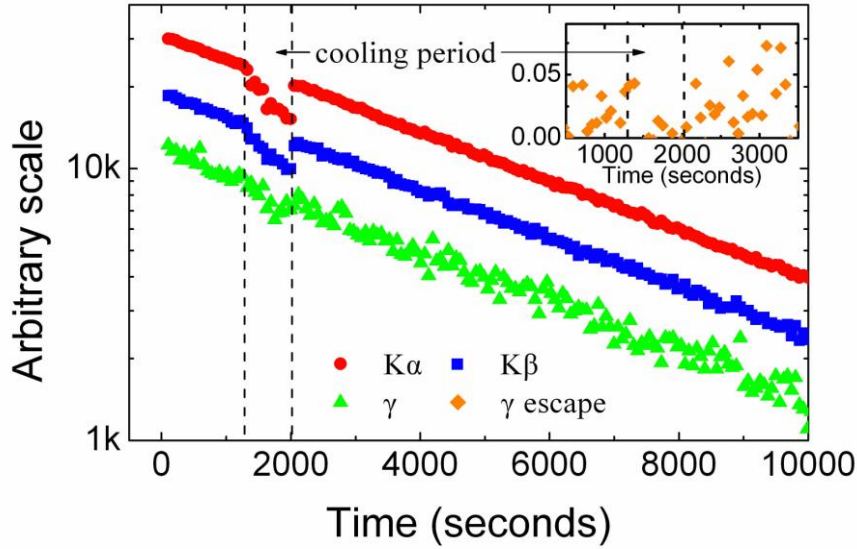

FIG. 3. (Color online) Time evolutions of two K lines, γ and γ escape emitted from the long-lived $^{103}$Rh isomeric state. Kα lines are collected from 13 channels between 19.4 keV and 20.7 keV, and Kβ lines are collected from 14 channels between 22.1 keV and 23.4 keV. γ is collected from 14 channels between 39.0 keV and 40.3 keV. The ordinate scale stands for the Kα but arbitrary for Kβ and γ. Superradiance channels open but not toward the detector, which is revealed by the recovery back to original decay lines immediately after cooling stop. Suppression of K lines and γ are attributed to directional superradiance of entangled cascade branching γs. Time evolution of 30-keV γ escape normalized by 40-keV γ is shown by the upper figure. Sag of γ escape induced by cooling reconfirms the superradiant cascade γs.

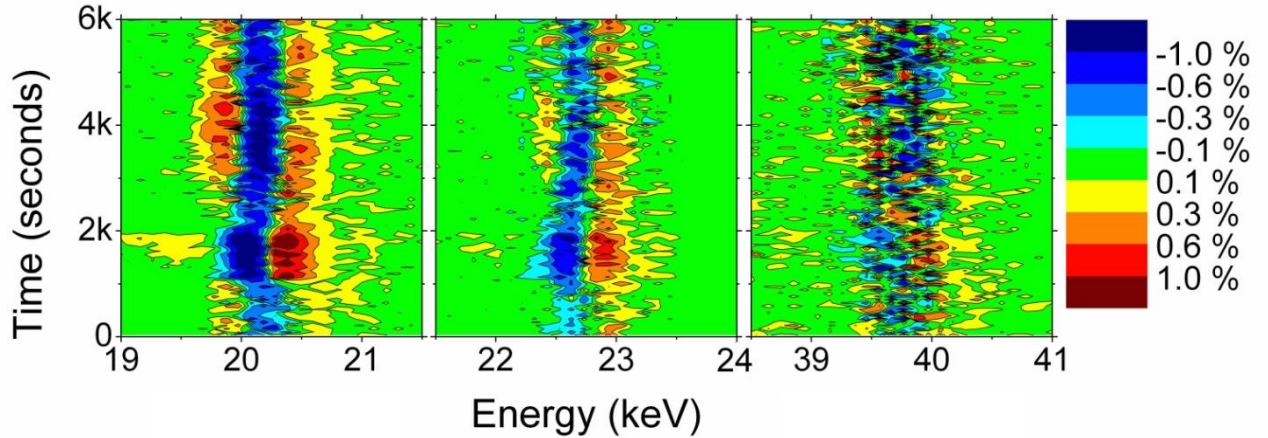

FIG. 4. (Color online) Time evolutions of three spectral deformations, i.e. Kα lines at 20 keV, Kβ lines at 23 keV, and γ at 40 keV. The baseline shift in the original data has been corrected. The cooling of liquid nitrogen starts at 1.2 ks and stops at 1.92 ks. Hypersatellites are increased by an order of magnitude after quenching. Total counts N(ω,t) in each band are normalized with N(Kα,t)=1, N(Kβ,t)=0.5 and N(γ,t)=0.25 for clear presentation as defined in eq. (1). $Kα_2$ is enhanced except during the cooling period, in which the satellite shift dominates. After cooling stop, hypersatellite shift increases in addition to satellite shift. Kβ hypersatellites are stronger than Kα hypersatellites. Spectral deformation of γ is true, which requires further investigation.

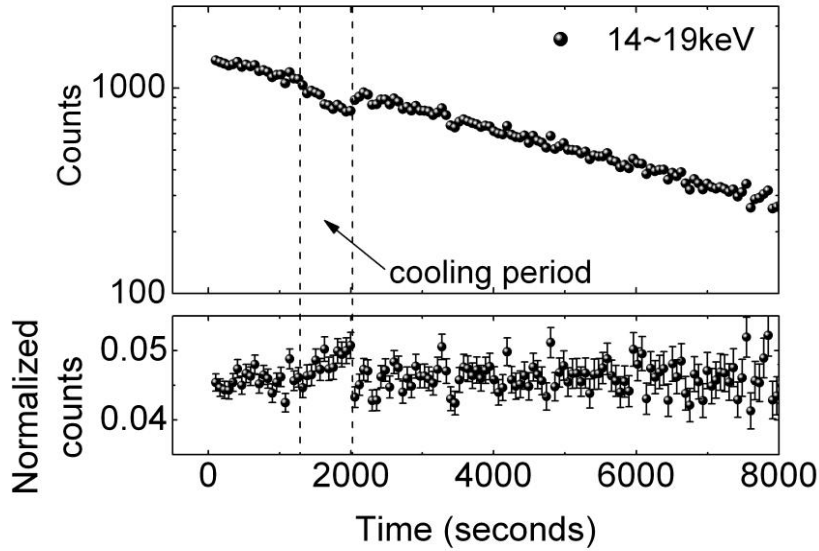

FIG. 5. Time evolution of cascade branching γs between 14 keV and 19 keV. Counts are normalized by Kα counts channel by channel in time. The level of 0.045 is mainly (70%) contributed by the detector imperfection induced by Kα counts at 20 keV. Elevation of the normalized counts during the cooling period reveals that the cascade branching γs follow the sag of 40-keV γ but increased by K suppression in Fig. 3.

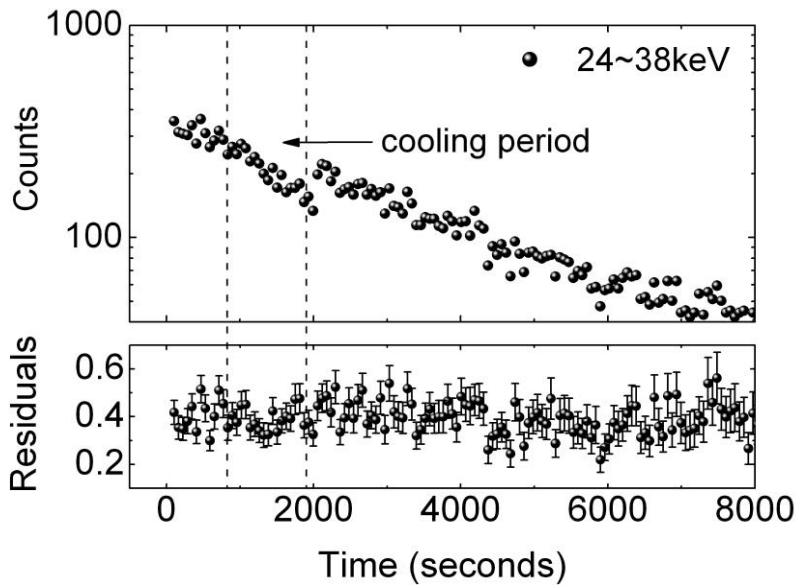

FIG. 6. Time evolution of cascade branching γs between 24 keV and 38 keV. The residuals are calculated by removing the tail pile-up of K lines and then normalized by the 40-keV γ counts channel by channel in time. The level of 0.4 represents the ratio between the cascade branching γs in this band and the 40-keV γ in Fig. 3, where counts of the Compton continuum are negligible. This level depends on collecting angle, eg. 0.4 for solid angle of $0.4\pi$ sr reported here, 0.3 for solid angle of $1.2\pi$ sr in [2] and 0.4 for solid angle of $1.7\pi$ sr in [1].

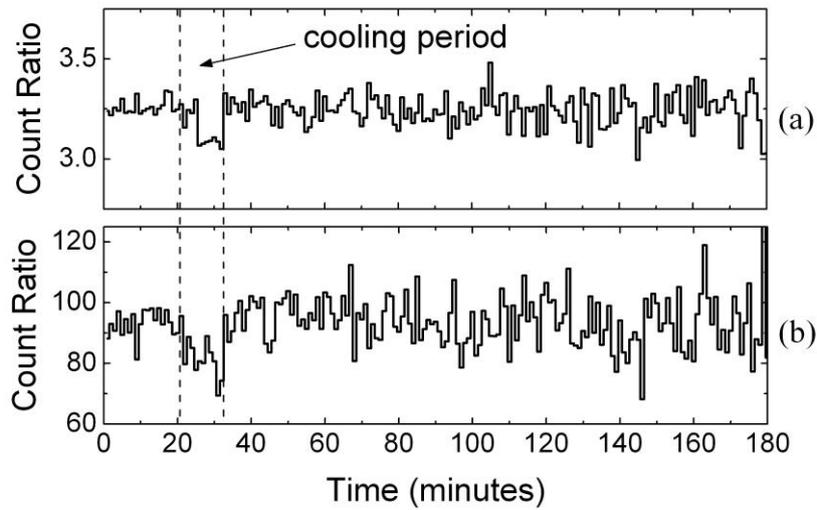

FIG. 7. Time evolutions of Kα/Kβ and K/γ ratio to demonstrate the negligible nitrogen absorption and stronger K suppression than γ suppression in Fig. 3. (a): In the first six minutes of cooling, most of the liquid nitrogen between sample and detector is evaporated. No significant nitrogen absorption is observed. (b): K/γ ratio reveals that the K suppression is stronger than γ suppression, particular during the first six minutes of cooling without any nitrogen absorption. After the cooling stop, the slow increment of K/γ reveals the speed-up γ and slow-down K decay reported in [1].